\documentclass[english,notitlepage,nofootinbib,superscriptaddress,tightenlines]{revtex4-2}
\usepackage[T1]{fontenc}
\usepackage[usenames,dvipsnames]{color}
\usepackage[normalem]{ulem}
\usepackage{ulem}
\usepackage[utf8]{inputenc}
\usepackage{textcomp}

\usepackage[T1]{fontenc}
\usepackage{ae,aecompl}
\usepackage[usenames,dvipsnames]{color}  
\usepackage{tikz}
\usetikzlibrary{chains,shapes,fit,calc}
\usetikzlibrary{positioning,arrows}
\usepackage{ulem}
\usepackage{amsfonts}

\usepackage{geometry}
\usepackage{tikz-feynman}
\tikzfeynmanset{compat=1.0.0}

\usetikzlibrary{patterns,decorations.pathreplacing}
\usetikzlibrary{decorations.pathmorphing}
\usetikzlibrary{decorations.markings}
\usetikzlibrary{arrows,shapes}
\usetikzlibrary{shapes.misc}
\tikzset{
  gluon/.style={decorate, draw=black,
    decoration={coil,amplitude=4pt, segment length=4pt,aspect=0.7}}
}
\tikzset{
  photon/.style={decorate, decoration={snake}},
}
\usepackage{adjustbox}
\usepackage{multirow}
\definecolor{darkred}{rgb}{0.6,0,0}
\definecolor{brown}{rgb}{0.59, 0.29, 0.0}
\usepackage{amsmath}
\usepackage{dsfont}
\usepackage{graphicx}
\usepackage{placeins}
\usepackage[colorlinks=true,citecolor=darkred,urlcolor=darkred, pdfborder={0 0 0}]{hyperref}

\newcommand {\ignore}[1]{}
\def\321{$\mathrm{SU(3) \otimes SU(2) \otimes U(1)}$ }

\usepackage{booktabs}
\usepackage{orcidlink}

\def\gsim{\raise0.3ex\hbox{$\;>$\kern-0.75em\raise-1.1ex\hbox{$\sim\;$}}}
\def\lsim{\raise0.3ex\hbox{$\;<$\kern-0.75em\raise-1.1ex\hbox{$\sim\;$}}}

\def\beqn#1{\begin{equation}\label{#1}}
\def\eeqn{\end{equation}}

\def\beqa#1{\begin{eqnarray}\label{#1}}
\def\eeqa{\end{eqnarray}}

\def\znbb {neutrinoless double beta decay }

\def\Z2{$\mathcal{Z_2}$}

\def\321{$\mathrm{SU(3) \otimes SU(2) \otimes U(1)}$ }

\newcommand{\AddrAHEP}{%
  AHEP Group, Institut de F\'{i}sica Corpuscular --
  CSIC/Universitat de Val\`{e}ncia, Parc Cient\'ific de Paterna.\\
 C/ Catedr\'atico Jos\'e Beltr\'an, 2 E-46980 Paterna (Valencia) - SPAIN}
 
 \newcommand{\AddrIITBHU}{ Department of Physics, Indian Institute of Technology (BHU), Varanasi 221005, India}

  \newcommand{\AddrIISERB}{Department of Physics,
 Indian Institute of Science Education and Research - Bhopal \\
 Bhopal Bypass Road, Bhauri, Bhopal, India}

\begin{document}

\bibliographystyle{unsrt}   

\title{\boldmath \color{BrickRed} Zooming in on `bi-large' neutrino mixing with the first JUNO results}

 \author{Gui-Jun Ding}\email[Email Address: ]{dinggj@ustc.edu.cn}
 \affiliation{Department of Modern Physics, and Anhui Center for fundamental sciences in theoretical physics, \\
 University of Science and Technology of China, Hefei, Anhui 230026, China}
 \author{Ranjeet Kumar}\email[Email Address: ] {ranjeet20@iiserb.ac.in}
 \affiliation{\AddrIISERB}
 \author{Newton Nath}
 \email[Email Address: ]{nnath.phy@iitbhu.ac.in}
 \affiliation{\AddrIITBHU}
 \author{Rahul Srivastava}
 \email[Email Address: ]{rahul@iiserb.ac.in}
 \affiliation{\AddrIISERB}
 \author{Jos\'{e} W. F. Valle}
 \email[Email Address: ]{valle@ific.uv.es}
 \affiliation{\AddrAHEP}

\begin{abstract}

\vspace{1.0cm}
{\noindent
The leptonic mixing matrix is examined within bi-large mixing patterns and confronted with the latest results announced by the Jiangmen Underground Neutrino Observatory (JUNO). We analyze the viability of bi-large mixing schemes and assess JUNO’s ability to test neutrino mixing and discriminate among different bi-large mixing patterns, some of which are strongly disfavored when compared with neutrino oscillation global-fit results. Specific octant and CP predictions emerge. Finally, we comment on the implications of JUNO’s findings for neutrinoless double beta decay.
}
\end{abstract}
\maketitle

\section{Introduction}


The next-generation Jiangmen Underground Neutrino Observatory (JUNO) has released its first measurements of two of the neutrino oscillation parameters, \(\Delta m^2_{21}\) and \(\sin^2\theta_{12}\), based on 59.1 days of data. The reported best-fit values with \(1\sigma\) uncertainties are~\cite{JUNO:2025gmd}
\begin{align}
\sin^2\theta_{12} &= 0.3092 \pm 0.0087\,, \quad 
\Delta m^2_{21} = (7.50 \pm 0.12) \times 10^{-5}~\text{eV}^2\,,
\end{align}
for the normal neutrino mass ordering scenario.
Recently, the SNO+ collaboration also reported new measurements of reactor antineutrino oscillations~\cite{SNO:2025chx}. When combined with the PDG 2025 global dataset \cite{ParticleDataGroup:2024cfk}, the updated best-fit oscillation parameters are
$\sin^2\theta_{12} = 0.310 \pm 0.012\,,\,\Delta m_{21}^2 = (7.63 \pm 0.17) \times 10^{-5}~\text{eV}^2~.$

Remarkably, the JUNO results already surpass the precision of current neutrino oscillation global-fit determinations~\cite{deSalas:2020pgw,10.5281/zenodo.4726908,Esteban:2024eli,Capozzi:2025wyn,Capozzi:2025ovi}.
This improved precision provides an important new input for testing theoretical frameworks of lepton mixing and significantly sharpens constraints on many flavour models~\cite{King:2017guk,Feruglio:2019ybq,Xing:2020ijf,Almumin:2022rml,Chauhan:2023faf,Ding:2023htn,Ding:2024ozt}.
At the current juncture, the unknowns in the neutrino oscillation sector include the neutrino mass ordering (i.e., the sign of $\Delta m^{2}_{31}$, with $\Delta m^{2}_{31} > 0$ referred to as normal ordering and $\Delta m^{2}_{31} < 0$ as inverted ordering), the octant of the atmospheric mixing angle $\theta_{23}$ (i.e., $\theta_{23} < 45^\circ$ for the lower octant and $\theta_{23} > 45^\circ$ for the upper octant), and the CP-violating phase $\delta$.   
Considerable theoretical and experimental efforts are currently underway to predict and determine these unknowns.
With the release of the first data from JUNO, there have already been several attempts to assess the impact of these precision measurements of oscillation parameters on existing theoretical frameworks~\cite{Chao:2025sao,Li:2025hye,Zhang:2025jnn,Huang:2025znh,Ge:2025cky,Xing:2025xte,Chen:2025afg,He:2025idv,Jiang:2025hvq,Petcov:2025aci,Ge:2025csr}. 
However, it becomes more challenging to develop theoretical models that describe the increasingly precise experimental data. Indeed, there have been several recent theoretical approaches to understand neutrino mixing and oscillations~\cite{King:2017guk,Feruglio:2019ybq,Xing:2020ijf,Almumin:2022rml,Chauhan:2023faf,Ding:2023htn,Ding:2024ozt}. 

Motivated by the latest precision measurements from JUNO, we examine the framework of bi-large neutrino mixing, as proposed in~\cite{Boucenna:2012xb,Ding:2012wh,Roy:2014nua}, which provides sharp predictions for the
neutrino mixing parameters. 
The idea behind such mixing patterns relies on the observation that the smallest lepton mixing angle is numerically close to the largest quark mixing angle. This leads to the hypothesis that the Cabibbo angle may serve as a universal seed for both quark and lepton mixings. Such a bi-large mixing paradigm has already inspired some approaches for describing the structure of the lepton mixing matrix ~\cite{Boucenna:2012xb,Ding:2012wh,Roy:2014nua}. 
Recently, more refined variants of this approach have been proposed~\cite{Chen:2019egu,Ding:2019vvi}, further motivating an updated analysis in light of the improved experimental precision reached in JUNO. 

The lepton mixing matrix is the leptonic analogue of the Cabibbo-Kobayashi-Maskawa (CKM) matrix. It is defined as 
\begin{equation}
 U = U^\dagger_{\ell}\, U_{\nu},   
\end{equation}
where \(U_{\ell}\) and \(U_{\nu}\) denote the diagonalization matrices associated with the charged-lepton and neutrino sectors, respectively.
The peculiar pattern of its matrix elements has intrigued theorists ever since neutrino oscillations were discovered. This interest has only intensified following reactor measurements confirming that \(\theta_{13} \neq 0\)~\cite{An:2012eh,An:2016ses} and T2K results suggesting a nearly maximal value of the leptonic CP-violating phase \(\delta\)~\cite{Abe:2017uxa}. 
Several attempts have been made to account for the non-zero value of \(\theta_{13}\)~\cite{Minakata:2004xt,Albright:2010ap,Chen:2015siy,Pasquini:2016kwk,Chen:2018lsv}, including deviations from the conventional tribimaximal mixing pattern~\cite{Harrison:2002er,Rahat:2018sgs}, which can yield realistic descriptions of neutrino oscillation data~\cite{Chen:2018eou,Chen:2018zbq}. 

In this work, we examine the compatibility of bi-large proposals for the leptonic mixing matrix by confronting them with the latest JUNO measurement as well as global analysis of neutrino oscillation data~\cite{deSalas:2020pgw,Esteban:2024eli,Capozzi:2025wyn}.  
In addition to the bi-large \textit{ansatz} for the neutrino diagonalization matrix $U_{\nu}$, these schemes include a charged-lepton correction factor $U_{\ell}$, taken to be ``CKM-like,'' as motivated by Grand Unified Theories (GUTs). 
Our approach follows the same general strategy as the symmetry-based sensitivity studies performed in the context of the DUNE experiment in Refs.~\cite{Chatterjee:2017ilf,Srivastava:2017sno,Srivastava:2018ser,Chakraborty:2018dew,Nath:2018xkz,Nath:2018fvw}.
Our present study is specifically tailored to address the impact of the recent JUNO data on the bi-large lepton mixing schemes.
We determine the allowed regions of the oscillation parameters within each bi-large \emph{ansatz} and discuss the prospects for distinguishing among the different bi-large scenarios.

\section{Bi-large Mixing Patterns}
\label{sec:BiLar}

Here, we present an overview of the predictions of bi-large mixing patterns for the lepton mixing matrix. For concreteness, we focus on the patterns T1 and T2 from Ref.~\cite{Chen:2019egu}, while T3 and T4 are taken from the earlier work in Ref.~\cite{Roy:2014nua}. Following Ref.~\cite{Roy:2014nua}, we assume that the bi-large patterns arise from simple Grand Unified Theory (GUT) constructions, in which the charged-lepton and down-type quark sectors exhibit similar structure.

\subsection*{Type-1 (T1)}


The first bi-large pattern, {\bf T1}, assumes that the neutrino mixing angles are correlated with the Cabibbo angle as follows~\cite{Chen:2019egu}
\begin{equation}
\sin\theta_{23} = 1 - \lambda, ~~\sin\theta_{12} = 2\lambda, ~~\sin\theta_{13} = \lambda \;.
  \end{equation}
 Truncating to $\mathcal{O}(\lambda^{2})$, we obtain the neutrino factor of the mixing matrix $U_{BL1}$ as 
\begin{align}
U_{BL1}\approx \left[
\begin{array}{ccc}
 1-\frac{5 \lambda ^2 }{2} &~ 2 \lambda &~ - \lambda  \\
\lambda - 2\sqrt{2} \lambda^{3/2}&~ \sqrt{2 \lambda} - \dfrac{\lambda^{3/2}}{2\sqrt{2}} &~ 1 - \lambda - \dfrac{\lambda^2}{2} \\
2 \lambda + \sqrt{2} \lambda^{3/2} &~ -1 + \lambda &~ \sqrt{2 \lambda} - \dfrac{\lambda^{3/2}}{2\sqrt{2}} 
\end{array}
\right]\;.
\end{align}
By itself, $U_{BL1}$ alone does not account for the full leptonic mixing matrix. To obtain a realistic mixing matrix consistent with current oscillation data, we must include the appropriate charged-lepton corrections to $U_{BL1}$. For this case, the $SO(10)$ GUT–motivated, CKM-like charged-lepton corrections are given by 
\begin{align}\label{eq:Ul1}
U_{l_1}&=R_{23}(\,\theta_{23}^{CKM}\,)\Phi R_{12}(\,\theta_{12}^{CKM}\,)\Phi^{\dagger}\approx \begin{bmatrix}
1-\frac{1}{2}\lambda^2 & \lambda\, e^{-i \phi} & 0 \\
-\lambda\, e^{i \phi} & 1-\frac{1}{2}\lambda^2 & A\lambda^2 \\
A \lambda^3 e^{i \phi} & -A \lambda^2 & 1
\end{bmatrix}.
\end{align} 
Here we use $\sin\theta^{\rm CKM}_{12} = \lambda$ and $\sin\theta^{\rm CKM}_{23} = A\lambda^{2}$, where $\lambda=0.22501\pm 0.00068$ and \ignore{$A = 0.836 \pm 0.015$} $A=0.826^{+0.016}_{-0.015}$ are the Wolfenstein parameters~\cite{ParticleDataGroup:2024cfk}\ignore{\cite{Tanabashi:2018oca}}. 
We also define $\Phi = \mathrm{diag}\{ e^{-i\phi/2},\, e^{i\phi/2},\, 1 \}$, with $\phi$ treated as a free parameter. 
The lepton mixing matrix for \textbf{T1} is then given by $U = U_{l_1}^{\dagger} U_{BL1}$, from which the mixing angles and the leptonic Jarlskog invariant $J_{CP}$ follow as 
\begin{equation}
\label{eq:BL1}
\begin{aligned}
\sin^{2}\theta_{13} & \approx 4\lambda^{2} (1-\lambda)\cos^2\dfrac{\phi}{2} \,, \\
\sin^{2}\theta_{12} & \approx 2\lambda^{2} (2-2\sqrt{2\lambda}\,\cos\phi+\lambda) \,, \\
\sin^{2}\theta_{23} &\approx (1-\lambda)^{2}-2\sqrt{2} A\lambda^{5/2}-2\lambda^{3}(1+2\cos\phi) \,, \\
J_{CP}  & \approx -2\left(\sqrt{2}+\sqrt{\lambda}\right)\lambda^{5/2} \sin\phi \,.
\end{aligned}
\end{equation}
Note that, besides the leading order term, there is a $-2\lambda^3\sin\phi$ contribution in $J_{CP}$.

\subsection*{Type-2 (T2)}

In the second bi-large pattern, \textbf{T2}, we relate the neutrino mixing angles to the Cabibbo angle as follows~\cite{Chen:2019egu},
\begin{equation}
\sin\theta_{23} = 3\lambda, ~~ \sin\theta_{12} = 2\lambda, 
~~\sin\theta_{13} = \lambda~~ 
\end{equation}
and approximate the neutrino diagonalization matrix as 
\begin{align}
U_{BL2}\approx \left[
\begin{array}{ccc}
 1-\frac{5 \lambda ^2 }{2} &~ 2 \lambda &~ - \lambda  \\
-2\lambda + 3\lambda^{2} &~ 1 - \dfrac{13 \lambda^{2}}{2} &~ 3\lambda \\
\lambda + 6 \lambda^{2} &~ -3\lambda + 2\lambda^{2} &~ 1 - 5\lambda^{2}\\
\end{array}
\right]
\end{align}
Motivated by $SU(5)$ unification, here we take the charged-lepton diagonalization matrix to be 
\begin{align}\label{eq:Ul2}
U_{l_2}& = \Phi^{\dagger} R^{T}_{12}(\,\theta_{12}^{CKM}\,)\Phi R^{T}_{23}(\,\theta_{23}^{CKM}\,)\approx \begin{bmatrix}
1-\frac{1}{2}\lambda^2 & -\lambda\, e^{i \phi} & A \lambda^3 e^{i \phi} \\
\lambda\, e^{- i \phi} & 1- \frac{1}{2}\lambda^2 & - A\lambda^2 \\
0 & A \lambda^2 & 1
\end{bmatrix}\,, 
\end{align}
where $\theta_{12}^{\rm CKM}$, $\theta_{23}^{\rm CKM}$, and $\Phi$ have the same definitions as given below Eq.~\eqref{eq:Ul1}. 
Using the forms of $U_{BL2}$ and $U_{l_2}$, we construct the lepton mixing matrix for \textbf{T2} as 
$U = U_{l_2}^{\dagger} U_{BL2}$. The resulting mixing angles and the leptonic Jarlskog invariant $J_{CP}$ are 
\begin{equation}
\label{eq:BL2}
\begin{aligned}
\sin^{2}\theta_{13} &\approx\lambda^{2}-6\lambda^3\cos\phi+ 8\lambda^4 \,, \\
\sin^{2}\theta_{12} & \approx\lambda^{2} (5+4\cos\phi)-2\lambda^4(8+13\cos\phi) \,, \\
\sin^{2}\theta_{23} & \approx 9\lambda^{2}+6\lambda^{3}(A+\cos\phi)-\lambda^4(8-2A\cos\phi-A^2)\,, \\
J_{CP} &\approx -\left[3+(16+A)\lambda\right]\lambda^{3} \sin\phi\,,
\end{aligned}
\end{equation}
where the next-to-leading-order contribution $-(16 + A)\lambda^{4}\sin\phi$ in $J_{CP}$ is included, as it is comparable in size to the leading-order term.


\subsection*{Type-3 (T3)}

%
The patterns {\bf T3} and {\bf T4} require, in addition to the standard Wolfenstein parameters $\lambda$ and $A$, two other free parameters $\psi$ and $\delta$ in order to fully characterize their physical lepton mixing matrices. 
To first order approximation, the three neutrino mixing angles are assumed to be of the following form, 
\begin{equation}
\label{eq:lepton-angles-BL34}\sin\theta_{12}=\sin\theta_{23} =\psi\lambda, \qquad \sin\theta_{13}=\lambda\,.
\end{equation}
where $\psi$ is a free parameter whose value is determined by fitting to neutrino oscillation data.  From Eq.~\eqref{eq:lepton-angles-BL34} it follows that the neutrino diagonalization matrix $U_{BL3}$ is given by
\begin{align}
 U_{BL3}=\left[
\begin{array}{ccc}
c\sqrt{1-\lambda^2}   ~&~   \psi\lambda\sqrt{1-\lambda^2}  ~&~  \lambda \\
-c\psi\lambda(1+\lambda)  ~&~   c^2-\psi^2\lambda^3   ~&~  \psi\lambda\sqrt{1-\lambda^2}  \\
-c^2\lambda+\psi^2\lambda^2  ~&~  -c\psi\lambda(1+\lambda)  ~&~  c\sqrt{1-\lambda^2}
\end{array}
\right],\quad c\equiv\cos \sin^{-1}(\psi\lambda)\,.
\label{eq:Ubl3}
\end{align}

For the case of \textbf{T3}, we take the charged-lepton diagonalization matrix to be the same as $U_{l_1}$ in Eq.~\eqref{eq:Ul1}. With this choice, we can obtain the following expressions for the lepton mixing parameters: 
\begin{equation}
\label{eq:bl3}
\begin{aligned}
\sin^2\theta_{13}&\approx\lambda ^2-2 \psi\lambda^3\cos\phi+\left(-1+2Ac\cos\phi+\psi^2\right)\lambda^4\,, \\
\sin^2\theta_{12}&\approx\left(c^4-2c^2\psi\cos\phi+\psi^2\right)\lambda^2+\left(c^4-\psi^2\right)\lambda^4\,, \\
\sin^2\theta_{23}&\approx\psi^2\lambda^2+2\left(\cos\phi-A c\right)\psi\lambda^3+\left(1-\psi^2-2Ac\cos\phi+A^2c^2\right)\lambda^4\,,\\
J_{CP}&\approx c^2\left[\psi+\left(\psi-Ac-\psi^3\right)\lambda\right]\lambda^3\sin\phi\,.
\end{aligned}
\end{equation}


\subsection*{Type-4 (T4)}

For this bi-large mixing pattern, \textbf{T4}, we again take the neutrino part of the mixing matrix to be identical with $U_{BL3}$, as given in Eq.~\eqref{eq:Ubl3}.  The distinction between the \textbf{T3} and \textbf{T4} patterns arises from the choice of charged-lepton corrections.  
In contrast to the \textbf{T3} pattern, which uses the matrix $U_{l_1}$, here we adopt the charged-lepton diagonalization matrix $U_{l_2}$ introduced in Eq.~\eqref{eq:Ul2}.  
This modification leads to different predictions for the lepton mixing parameters.
In this setup, the lepton mixing angles and the Jarlskog invariant are determined to be: 
\begin{equation}
\label{eq:bl4}
\begin{aligned}
\sin^{2}\theta_{13} &\approx \lambda^{2}+2\,\psi\,\lambda^{3}\cos\phi-\left(1-\psi^{2}\right)\lambda^{4}\,, \\[4pt]
\sin^{2}\theta_{12} &\approx\left(c^{4}+2c^{2}\psi\cos\phi+\psi^{2}\right)\lambda^{2}
+\left(c^{4}-\psi^{2}\right)\lambda^{4}\,, \\[4pt]
\sin^{2}\theta_{23} &\approx \psi^{2}\lambda^{2}+2\left(Ac-\cos\phi\right)\psi\lambda^{3}+\left(1-\psi^{2}-2Ac\cos\phi+A^{2}c^{2}\right)\lambda^{4}\,, \\
J_{CP} &\approx c^{2}\left[\psi+\left(\psi+Ac-\psi^{3}\right)\lambda\right]\lambda^{3}\sin\phi \;.
\end{aligned}
\end{equation}

Having introduced the full set of bi-large mixing patterns, including their corresponding charged-lepton corrections, we now proceed in the next section to examine their phenomenological implications in light of recent JUNO results and other current neutrino oscillation data

\section{Status of Bi-Large Mixing before JUNO's $\theta_{12}$ measurement}
\label{sec:global-fit}

The predictions of the four bi-large mixing patterns discussed here can be tested in a number of neutrino oscillation experiments. Before exploring the impact of recent JUNO results, let us briefly investigate their status before the inclusion of JUNO data, updating the study given in~\cite{Ding:2019vvi}. 
The status of the bi-large mixing patterns can be nicely encoded through the $(\sin^2\theta_{23} - \delta)$, $(\sin^2\theta_{13} - \delta)$, and $(\sin^2\theta_{13} - \sin^2\theta_{23})$ oscillation parameter correlation plots.

\begin{figure}[h!t]
    \centering
     \hspace{-1.0cm}
     \includegraphics[width=0.32\linewidth]{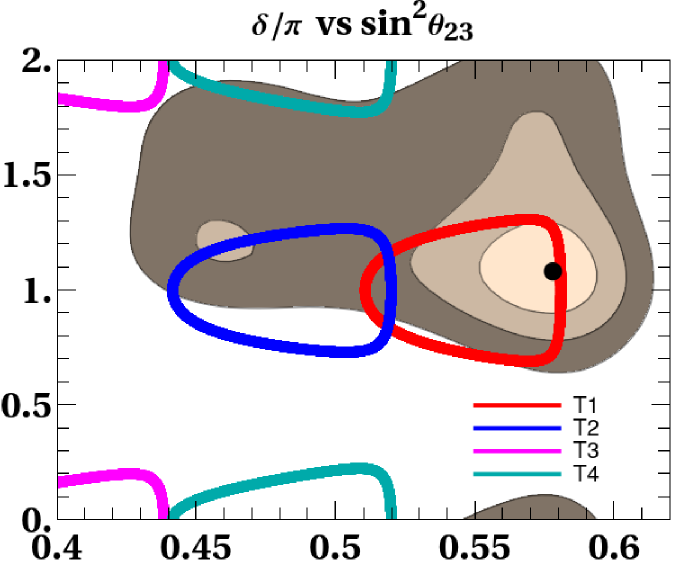}
     \includegraphics[width=0.33\linewidth]{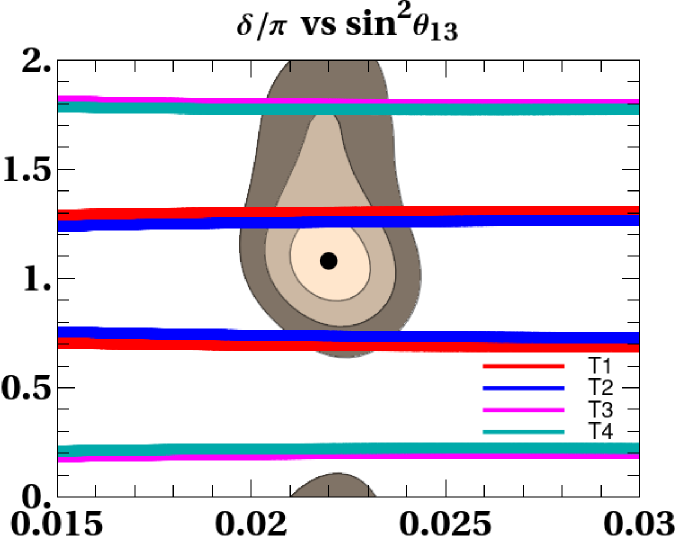}
     \includegraphics[width=0.34\linewidth]{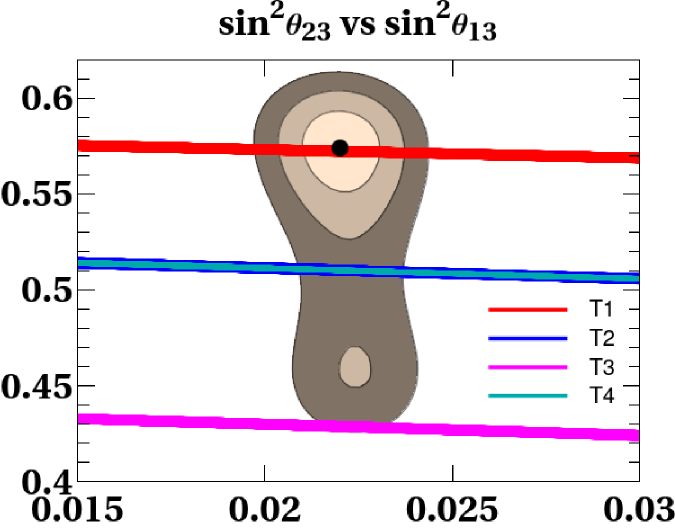}     
    \caption{\footnotesize
    Status of bi-large mixing pattern given the global oscillation fit in~\cite{deSalas:2020pgw,10.5281/zenodo.4726908}. The red, blue, magenta, and cyan 
    branches correspond to the \textbf{T1}, \textbf{T2}, \textbf{T3}, and \textbf{T4} patterns, respectively.}
    \label{fig:Status}
\end{figure}

We examined the status of bi-large mixing schemes taking into account the global fit of the world's oscillation data sample presented in~\cite{deSalas:2020pgw}. The results are given in Fig.~\ref{fig:Status}. Gray contours represent the $3\sigma$ allowed global fit ranges, prior to the inclusion of the solar mixing angle $\sin^2{\theta_{12}}$ announced by JUNO.
The model predictions are shown in red, blue, magenta, and cyan corresponding to \textbf{T1}, \textbf{T2}, \textbf{T3}, and \textbf{T4}, respectively. Contours are obtained by varying the oscillation parameters within their $3\sigma$ ranges for two degrees of freedom. 
The black dot denotes the corresponding best-fit value.
Among all four patterns, \textbf{T1} remains the most compatible with the oscillation data, successfully describing the best-fit values in the $(\sin^2\theta_{23} - \delta)$ plane as well as $(\sin^2\theta_{13} - \sin^2\theta_{23})$ plane. The \textbf{T2} pattern is consistent at the $2\sigma$ confidence level in the $(\sin^2\theta_{23} - \delta)$ plane, at the lower octant region. In contrast, \textbf{T3} is excluded by the current global-fit at the $3\sigma$ confidence level (see the magenta curve), while \textbf{T4} falls barely inside it. With improved precision in the measurement of $\delta$, the \textbf{T4} pattern is expected to be ruled out in the near future.

Having analyzed the bi-large mixing patterns through the $(\sin^2\theta_{23} - \delta)$, $(\sin^2\theta_{13} - \delta)$, and $(\sin^2\theta_{13} - \sin^2\theta_{23})$ correlation plots, next we present the status of the solar mixing angle $\sin^2{\theta_{12}}$ before and after the JUNO measurements.

\section{Status of Bi-large mixing patterns after JUNO's precision measurement}
\label{sec:test}

The Jiangmen Underground Neutrino Observatory (JUNO) is a multipurpose, large-volume liquid-scintillator neutrino experiment designed primarily to determine the neutrino mass ordering with high statistical significance. 
The detector is located at a baseline of about 53~km from the Yangjiang and Taishan nuclear power plants, optimized to resolve oscillation effects associated with both the solar and atmospheric mass-squared differences ~\cite{deSalas:2020pgw,Esteban:2024eli,Capozzi:2025wyn}. 
JUNO consists of a 20~kton central detector filled with linear alkylbenzene liquid scintillator, viewed by approximately 18{,}000 high-quantum-efficiency 20-inch PMTs together with an additional system of about 25{,}000 3-inch PMTs, yielding an unprecedented energy resolution of roughly $3\%/\sqrt{E(\mathrm{MeV})}$ \cite{JUNO:2025gmd}. 
The experiment aims to achieve sub-percent precision on the solar mixing parameters $\theta_{12}$, $\Delta m_{21}^2$, and $|\Delta m_{31}^2|$, as well as to measure the reactor antineutrino spectrum with high accuracy, with the potential to probe new physics scenarios such as nonstandard interactions and sterile neutrinos. 

Before discussing the impact of the precision measurement of JUNO, we would like to point out that \textbf{T1} and \textbf{T2} are one-parameter patterns for the lepton mixing matrix, depending only on the parameter $\phi$, which is adjusted to reproduce all three mixing angles and the Dirac CP phase $\delta$. Consequently, the mixing angles and $\delta$ are strongly correlated. For a detailed discussion, see Ref.~\cite{Ding:2019vvi}. 
The allowed ranges of the model parameter $\phi \in (0.7502 \pi - 0.7767 \pi)$ are obtained through the random scan by requiring all mixing angles and the Dirac phase $\delta$ to lie within their current $3\sigma$ experimental ranges for \textbf{T1}~\footnote{For completeness, we have performed the $\chi^2$ analysis to obtain the ranges for $\phi$. The $\chi^2$-function is given by
$
    \chi^2(\phi) = \sum_i \frac{(x_i(\phi) - \bar{x}_i)^2}{\sigma_i^2}, 
$
where $x_i(\phi) = (\sin^2\theta_{12}, \sin^2\theta_{13}, \sin^2\theta_{23}, \delta)$, and  $\bar{x}_i$ and $\sigma_i$ are the corresponding best-fit values and 1$\sigma$ uncertainties obtained in the global analysis of neutrino oscillation data.
Notice that the obtained values of $\phi$ at 3$\sigma$ C.L., for two degrees of freedom, remain very close to those found from the random scan.}~\cite{deSalas:2020pgw}.
By contrast, \textbf{T3} and \textbf{T4} depend on two parameters, $\psi$ and $\phi$.
In particular, the value of $\psi$ is restricted to a narrow band around $\psi = 3$ as discussed in Ref.~\cite{Ding:2019vvi}.
In what follows, we present the results of our numerical analysis for the various bi-large mixing schemes. Note that to obtain these we employ the exact formulae, rather than the leading-order approximations introduced in section~\ref{sec:BiLar}.

\begin{figure}[!t]
    \centering
    \hspace{-1.0cm}
    \includegraphics[width=0.33\linewidth]{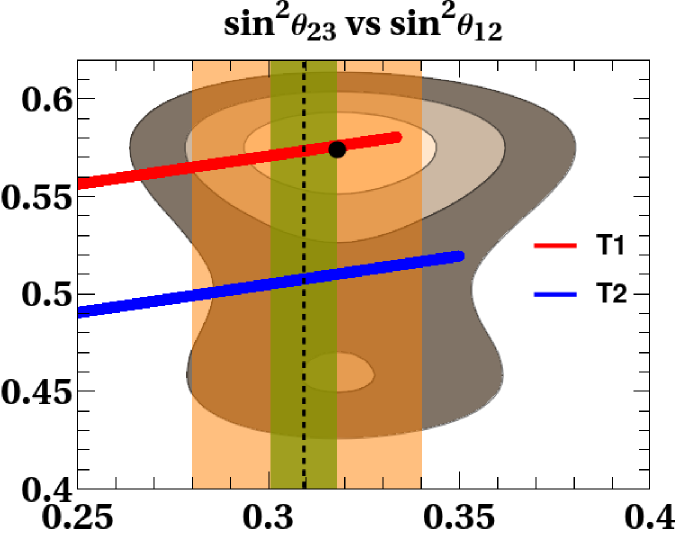} 
    \includegraphics[width=0.32\linewidth]{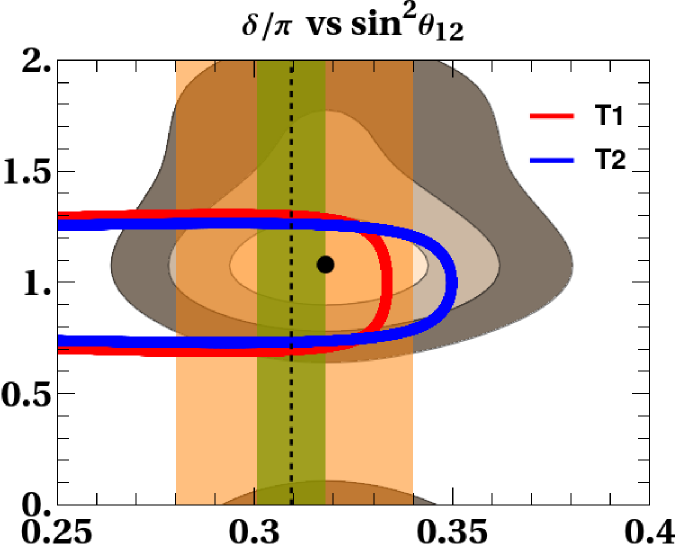}
     \includegraphics[width=0.34\linewidth]{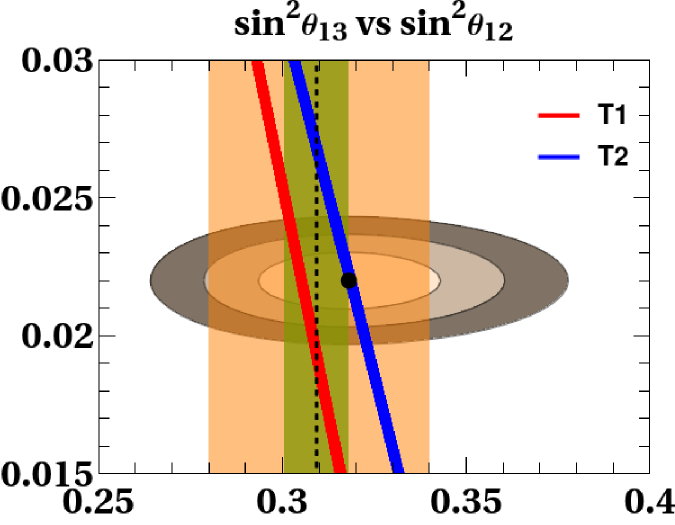} 
    \caption{\footnotesize  {
    The dashed black line gives the best fit value of $\sin^2\theta_{12}$ from JUNO with the corresponding $1\sigma$, $3\sigma$ ranges shown in green and orange, respectively. The allowed $3\sigma$ C.L. global-fit parameters are shown in dark gray, with the $1\sigma$, $2\sigma$ regions also displayed and the global best-fit marked (black dot). JUNO's potential to test the \textbf{T1}/\textbf{T2} bi-large schemes is seen by comparing the red/blue regions with JUNO and global-fit results. See text for details.
    }}
    \label{fig:T1}
\end{figure}

In the three panels of Fig.~\ref{fig:T1}, we show our results for the \textbf{T1} and \textbf{T2} mixing patterns in the 
$(\sin^2\theta_{12} - \sin^2\theta_{23})$ , $(\sin^2\theta_{12} - \delta)$, and $(\sin^2\theta_{12} - \sin^2\theta_{13})$
parameter spaces, respectively. 
In each panel, the \textbf{T1} predictions are shown in the red regions, while \textbf{T2} predictions are indicated in blue. 
These are obtained by varying the oscillation parameters within their $3\sigma$ ranges for two degrees of freedom.
The oscillation parameters allowed at the $3\sigma$ confidence level (C.L.) from the global-fit in~\cite{deSalas:2020pgw} are shown in dark gray, with the corresponding $1\sigma$ and $2\sigma$ regions also displayed. 
The global best-fit value is marked by the black dot. 
The JUNO best-fit value is indicated by the dotted black line, and its associated $1\sigma$ and $3\sigma$ intervals are represented by the green and orange bands, respectively.
The comparison of the blue and red regions with JUNO and global-fit results indicates the capability of JUNO to test the \textbf{T1} and \textbf{T2} bi-large mixing proposals.

The first remarkable result is seen from the first panel, namely that the \textbf{T1} pattern is compatible only with the higher atmospheric octant  ($\sin^2\theta_{23} > 0.5$), thereby excluding maximal atmospheric mixing  ($\sin^2\theta_{23} = 0.5$). Including the JUNO result further tightens the prediction to a narrow range around $\sin^2\theta_{23} \simeq 0.57$. In contrast, as seen from the figure, we find that the \textbf{T2} pattern prefers nearly maximal mixing, $\sin^2\theta_{23} \approx 0.5$ at $3\sigma$.

The middle panel shows a very stringent restriction on the allowed Dirac CP phase $\delta$. Imposing the latest JUNO constraint on $\sin^2\theta_{12}$ splits the bi-large CP prediction into two branches, yielding  $\delta \approx 0.7\pi$ and $1.3\pi$ for the \textbf{T1} pattern at $1\sigma$. This outcome disfavors both maximal CP violation ($\delta = \pm \pi/2$) as well as
CP conservation ($\delta = \pi$) at $1\sigma$.
However, the latter remains allowed at $3\sigma$, as indicated by the red branch falling within the JUNO band (orange). 
In contrast, we note that the \textbf{T2} mixing pattern excludes CP conservation even at $3\sigma$, since in this region of $\delta$ the blue branch lies outside the orange band.

Concerning the solar mixing parameter, from the first two panels, one sees that the \textbf{T1} pattern yields  
$\sin^2\theta_{12} \sim (0.27 - 0.33)$, consistent with the global oscillation fit at $3\sigma$.  However, since $\theta_{13}$ is the most precisely measured mixing angle, the $(\sin^2\theta_{12}\text{ vs. }\sin^2\theta_{13})$ plane (third panel) gives a sharper restriction on the allowed region, yielding 
$\sin^2\theta_{12} \sim (0.30 - 0.31)$.  
Likewise, the \textbf{T2} pattern limits $\sin^2\theta_{12}$ to $\sim (0.315 - 0.325)$, as shown by the blue curve in the third panel.

\begin{figure}[!h]
    \centering
     \hspace{-1.0cm}
    \includegraphics[width=0.33\linewidth]{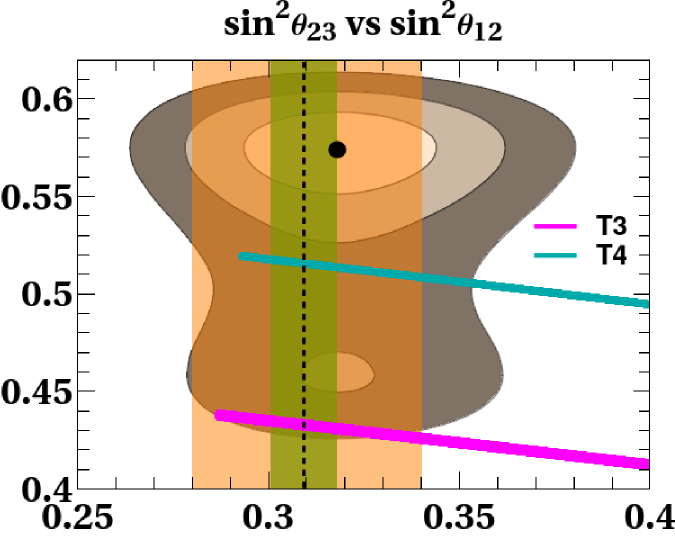}
    \includegraphics[width=0.32\linewidth]{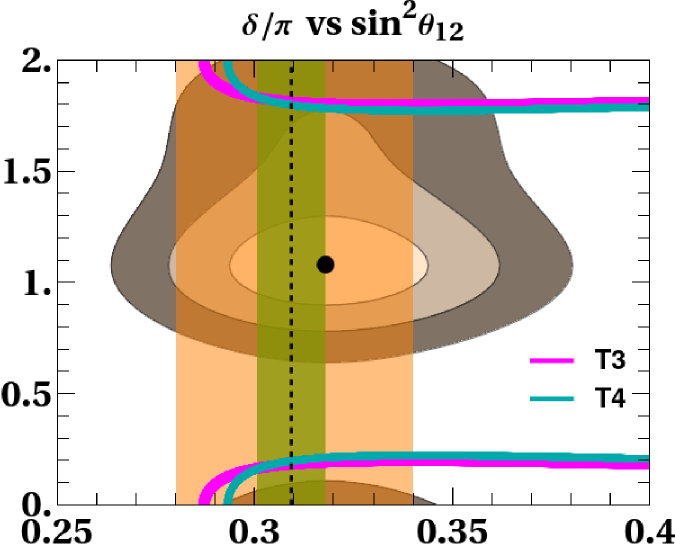}
    \includegraphics[width=0.34\linewidth]{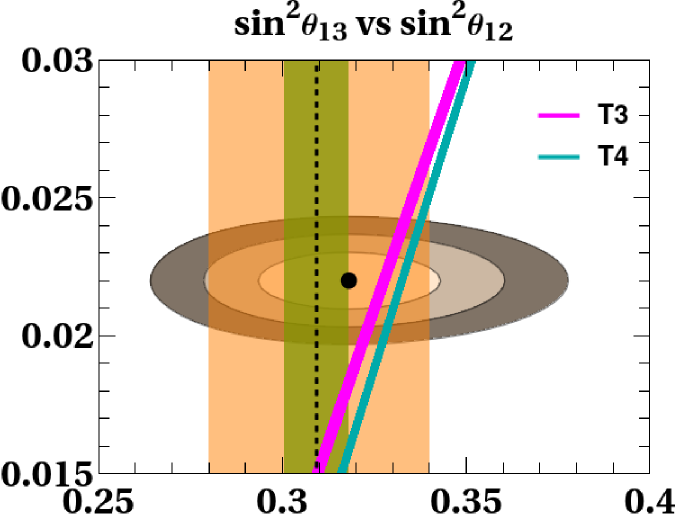}
    \caption{\footnotesize Same as Fig.~\ref{fig:T1}, but for type {\bf T3} (magenta color) and {\bf T4} (cyan color).}
    \label{fig:T4}
\end{figure}

The predictions for the \textbf{T3} and \textbf{T4} patterns are shown in Fig.~\ref{fig:T4}. 
Similar to the earlier plots, these results are also obtained by varying the remaining parameters within their respective $3\sigma$ ranges for two degrees of freedom.
From the first panel, we see that the \textbf{T3} pattern is only marginally compatible with the global-fit range of $\sin^2\theta_{23}$ at the $3\sigma$ level, and it predicts exclusively the lower octant (i.e., $\sin^2\theta_{23} < 0.5$). In contrast, the \textbf{T4} pattern predicts an atmospheric mixing angle that is close to maximal. 

The middle panel shows that both the \textbf{T3} and \textbf{T4} patterns are excluded at the $2\sigma$ confidence level. However, at $3\sigma$, one of the branches remains consistent with the global-fit. When the latest $1\sigma$ range from JUNO is included, the allowed region constrains the CP phase to lie near $1.8\pi$, thereby excluding CP conservation. At the $3\sigma$ level, however, compatibility with CP conservation (i.e., $\delta = 2\pi$) is recovered.
From the third panel, we see that while both patterns remain compatible with the global-fit to oscillation data at the $1\sigma$ level, they become excluded when the JUNO $1\sigma$ constraint is applied.
However, at the $3\sigma$ level, both models are compatible with JUNO data.
\textit{Overall, the \textbf{T3} pattern is expected to be decisively tested---and potentially ruled out---by future precision measurements of $\sin^2\theta_{23}$ and $\delta$.}

\section{Neutrinoless double beta decay after JUNO's precision measurement}

\begin{figure}[h!]
    \centering
     \hspace{-1.0cm}
     \includegraphics[width=0.52\linewidth]{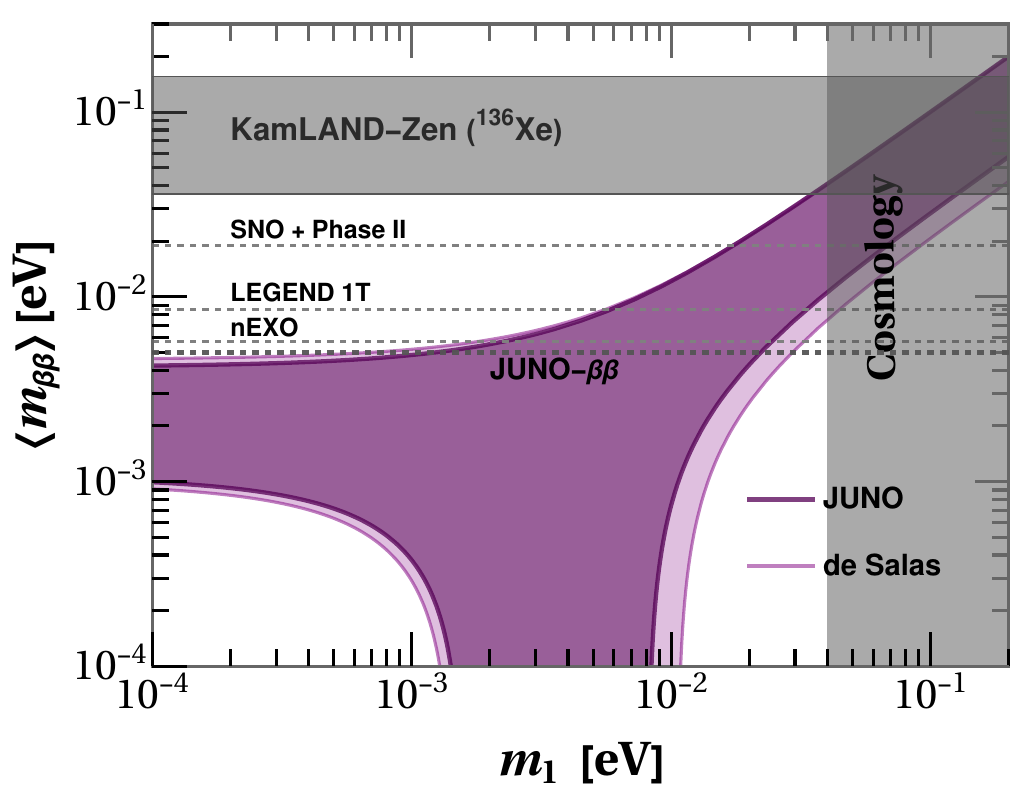}
     \includegraphics[width=0.52\linewidth]{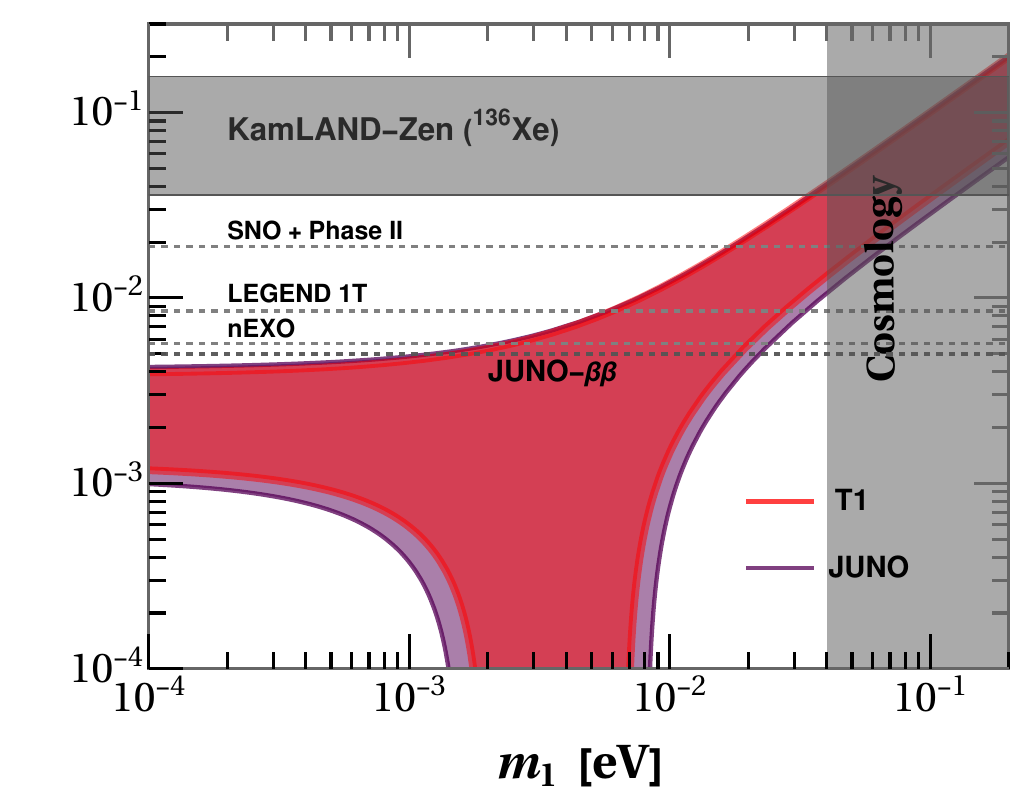}
    \caption{\footnotesize 
    Effective Majorana neutrino mass parameter $\langle m_{\beta \beta}\rangle$ as a
function of the lightest active neutrino mass $m_1$ for normal ordering. The left panel compares the allowed $\langle m_{\beta \beta}\rangle$ region for a generic scheme before and after JUNO's results, shown in light and dark purple, respectively. The right panel compares
the {\bf T1} bi-large pattern (in red) with the JUNO results (dark purple).
The current KamLAND-Zen limit and the projected sensitivities, including JUNO, are also indicated. The limit on $m_1$ from cosmology is shown in vertical gray band; see text for details.}
    \label{fig:mee}
\end{figure}

We now turn to the predictions for neutrinoless double beta ($0\nu\beta\beta$) decay within this framework. These also involve the Majorana phases, inaccessible to oscillation experiments, though oscillations leave an imprint on the allowed values of the corresponding decay amplitude, determined by the effective Majorana mass parameter $\langle m_{\beta \beta} \rangle$. In the symmetrical parametrization, it is given by \cite{Rodejohann:2011vc}
\begin{align}
    \langle m_{\beta \beta} \rangle = \left | \sum_i U^2_{ei} m_i \right| \equiv \left | c^2_{12} c^2_{13} m_1 + s^2_{12} c^2_{13} m_2 e^{2 i \phi_{12}} +  s^2_{13} m_3 e^{2 i \phi_{13}} \right |,
\end{align}
where $c_{ij}=\cos{\theta_{ij}}$, $s_{ij}=\sin{\theta_{ij}}$, $m_i$ are neutrino masses, and $\phi_{12}$, $\phi_{13}$ are the two physical Majorana phases~\cite{Schechter:1980gr,Schechter:1980gk}.
In the two panels of Fig.~\ref{fig:mee} we present $\langle m_{\beta \beta} \rangle$ as a function of the lightest neutrino mass $m_{1}$ for normal neutrino mass ordering, with the left panel corresponding to a generic scheme, while the right panel describes the $0\nu\beta\beta$ predictions for the \textbf{T1} bi-large mixing pattern in red and compares them with JUNO restrictions.
The left panel of Fig.~\ref{fig:mee} shows how the region allowed by the global oscillation fit (light purple) gets reduced by using the current JUNO results (dark purple). 

The right panel demonstrates that the bi-large pattern \textbf{T1} (red color region) fills a smaller region than allowed by the JUNO data. 
This is evident from the third panel of Fig.~\ref{fig:T1} that, when $\sin^2\theta_{13}$ is taken within the $3\sigma$ range, the allowed values of $\theta_{12}$ for the \textbf{T1} pattern are more constrained than the JUNO results at the $3\sigma$ C.L.
As the bi-large patterns do not predict mass-squared differences nor Majorana phases, all patterns \textbf{T1},  \textbf{T2}, \textbf{T3}, and \textbf{T4} have similar \znbb predictions. Thus, here we focus our discussion on the \textbf{T1} case only.

All in all, one sees that there is a slight reduction in the allowed \znbb region, stemming from JUNO's more precise measurements of $\theta_{12}$ and $\Delta m^{2}_{21}$.
As a visual guide, we also show current experimental limits from KamLAND-Zen (36–156 meV)~\cite{KamLAND-Zen:2022tow} as a horizontal gray band and
the projected sensitivities of next-generation experiments targeting 
$\langle m_{\beta\beta} \rangle \sim \mathcal{O}(10~\text{meV})$ at 90\% C.L.. These include 
JUNO-$\beta\beta$ $(5 - 12~\text{meV})$~\cite{Zhao:2016brs}, 
KamLAND2-Zen $(< 20~\text{meV})$~\cite{Nakamura:2020szx}, 
LEGEND~1000 $(8.5 - 19.4~\text{meV})$~\cite{LEGEND:2021bnm}, 
nEXO $(5.7 - 17.7~\text{meV})$~\cite{Albert:2017hjq}, 
and SNO$+$ Phase~II $(19 - 46~\text{meV})$~\cite{Andringa:2015tza}. 
These experiments are expected to significantly improve the exploration of the allowed value of $\langle m_{\beta\beta} \rangle $, although a substantial portion of the normal-ordering region will remain beyond their reach. Their expected sensitivities are shown in both panels of the figure, which also displays the neutrino mass constraint from cosmological observations by Planck~\cite{Aghanim:2018eyx} as a gray vertical band. The latest KATRIN result, $m_\beta < 0.45$ eV at $90\%$ C.L.~\cite{KATRIN:2024cdt}, can be approximately translated into the bound $m_1 < 0.45$ eV by considering the $3\sigma$ ranges of the relevant oscillation parameters, weaker than the cosmological constraint inferred from Planck~\cite{Aghanim:2018eyx}.

\FloatBarrier
\section{Summary}
\label{sec:summary}

We have described four bi-large patterns for the lepton mixing matrix.
We examined the compatibility of these four mixing patterns by confronting their predictions with current oscillation data as well as the latest JUNO measurement. 
First, we investigated their viability with global oscillation fits before JUNO, Fig.~\ref{fig:Status}.
By combining with the new JUNO constraint on $\sin^2\theta_{12}$ the bi-large mixing predictions become cornered by stringent restrictions on the allowed parameter space, see Figs.~\ref{fig:T1} and \ref{fig:T4}. 
For the \textbf{T1} pattern in particular, several distinctive features emerge: it is incompatible with maximal CP violation, it rules out the possibility of CP conservation, and it forbids maximal atmospheric mixing. These characteristics make its predictions sharply distinctive. On the other hand, the \textbf{T2} pattern predicts nearly maximal $\sin^2\theta_{23}$ and, concerning the CP phase, it
exhibits predictive behavior similar to that of the \textbf{T1} pattern.
Note that \textbf{T1} and \textbf{T2} are one-parameter schemes, characterized by a single parameter $\phi$ that determines all leptonic mixing angles and the Dirac CP phase $\delta$.
In contrast, the patterns \textbf{T3} and \textbf{T4} depend on two parameters, $\psi$ and $\phi$.
However, these patterns are severely cornered by the already existing data and are expected to be decisively tested and potentially excluded by future precision measurements of $\sin^2\theta_{23}$ and $\delta$.
We have also discussed the implications of JUNO’s findings for neutrinoless double beta decay, Fig.~\ref{fig:mee}.

\acknowledgments

This work is funded by Spanish grants PID2023-147306NB-I00 and Severo Ochoa Excellence grant CEX2023-001292-S (AEI/10.13039/501100011033) and by Prometeo CIPROM/2021/054 of Generalitat Valenciana. GJD is supported by the National Natural Science Foundation of China under Grant No~12375104 and Guizhou Provincial Major Scientific and Technological Program XKBF (2025)010.\\


\vspace{-4mm}

\FloatBarrier
\bibliographystyle{utphys}
\bibliography{bibliography}
\end{document}